\documentclass[aps,prl,superscriptaddress,preprintnumbers,showpacs,twoside,twocolumn,floatfix]{revtex4}

\usepackage{bm}
\usepackage{graphicx}
\usepackage{amsmath}

\usepackage{amsmath, amssymb}
\usepackage{times}
\usepackage{mathtools}

\begin{document}

\title{Order-by-disorder and quantum Coulomb phase in quantum square ice}

\author{Louis-Paul Henry}
\affiliation{Laboratoire de Physique, CNRS UMR 5672, Ecole Normale Sup\'erieure de Lyon, Universit\'e de Lyon, 46 All\'ee d'Italie, 
Lyon, F-69364, France}
\author{Tommaso Roscilde}
\affiliation{Laboratoire de Physique, CNRS UMR 5672, Ecole Normale Sup\'erieure de Lyon, Universit\'e de Lyon, 46 All\'ee d'Italie, 
Lyon, F-69364, France}

\begin{abstract}
We reconstruct the equilibrium phase diagram of quantum square ice, realized by the transverse-field Ising model on the checkerboard lattice, using a combination of quantum Monte Carlo, degenerate perturbation theory and gauge mean-field theory. The extensive ground-state degeneracy of classical square ice is lifted by the transverse field, leading to \emph{two} distinct order-by-disorder phases, a plaquette valence-bond solid for low field, and a canted N\'eel state for stronger fields. These two states appear via a highly non-linear effect of quantum fluctuations, and they can be identified with the phases of a lattice gauge theory (quantum link model) emerging as the effective Hamiltonian of the system within degenerate perturbation theory up to the 8th order. The plaquette valence-bond solid melts at a very low temperature, above which the system displays a thermally induced quantum Coulomb phase, supporting deconfined spinons.    
\end{abstract}



\maketitle

 \emph{Introduction.} Kinematically constrained systems represent a central theme of statistical physics and condensed matter, as they often represent the effective low-energy description of fundamental lattice many-body Hamiltonians. Prominent examples are to be found in models of frustrated magnetism (\emph{e.g.} frustrated Ising models \cite{MoessnerS2001}, quantum dimer models \cite{MoessnerR2011}) and of ice physics and its magnetic (spin-ice) analogs \cite{Castelnovoetal2012}. In such models the energy is typically minimized by an exponentially degenerate manifold of states satisfying a local constraint - the so-called ice rule. A fundamental insight is gained when recognizing that the ice rule can be cast in the form of a Gauss law for an emergent electric field -- corresponding, in the case of spin models, to the orientation of one of the spin components. When these systems are endowed with quantum dynamics, their effective Hamiltonian describing quantum fluctuations within the constrained manifold takes the form of a quantum \emph{lattice gauge theory} (LGT). A deconfined phase of the LGT - namely a phase in which the gauge field is not able to bind charges - corresponds to a novel phase supporting fractionalized excitations in the original spin model. Important examples thereof are represented by the deconfined phase of the Z$_2$ LGT in dimensions $d=2,3$ \cite{Kogut1979}, corresponding to the so-called Z$_2$ spin liquid in the magnetic context; and the deconfined phase of the $d=3$ compact lattice quantum electrodynamics (QED) \cite{Kogut1983}, corresponding to the so-called U(1) (or Coulomb) spin liquid \cite{Balents2010}. The former represents a strong candidate for the ground-state of frustrated $S=1/2$ Heisenberg antiferromagnets (\emph{e.g.} on the Kagom\'e lattice \cite{Yanetal2011}) while the latter is expected to be realized as the ground state of 3$d$ quantum spin ice \cite{Hermeleetal2004}. 
 
  In this context, a special role is played by two-dimensional (2$d$) quantum spin-ice models \cite{MoessnerS2001, Shannonetal2004, CastroNetoetal2006}. 2$d$ spin ice, or square ice, corresponds to the antiferromagnetic Ising model on the checkerboard lattice; its ground-state physics maps onto the 6-vertex model, whose phase space can be enumerated exactly \cite{Lieb1967}. The ensemble of the ice-rule states realizes a 2$d$ Coulomb phase \cite{Henley2010}, characterized by algebraic spin-spin correlations (decaying as $r^{-2}$) \cite{Sutherland1968} with a peculiar signature in the spin structure factor in the form of \emph{pinch points} \cite{Youngbloodetal1980}, and with \emph{deconfined} monopole-like excitations. When introducing quantum fluctuations in the system (either via a transverse-field term or via the coupling between the transverse spin components), a perturbative treatment of the quantum term to the lowest order leads to a model of \emph{frustrated} compact lattice QED (fcQED) for a discrete ($S=1/2$) gauge field \cite{CastroNetoetal2006,Hermeleetal2004} - also known as U(1) quantum link model or U(1) gauge magnet \cite{ChandrasekharanW1997}. Such a model can be suspected to undergo confinement due to the Polyakov mechanism valid for ordinary (non-frustrated) compact QED in $d=2$ \cite{Polyakovbook}; this implies that quantum effects remove the (deconfined) Coulomb phase through an order-by-disorder phenomenon, leading to a gapped ground state. This prediction is indeed consistent with numerics, finding a non-magnetic, gapped plaquette valence-bond solid (pVBS) as the ground state of fcQED \cite{Shannonetal2004, SyljuasenC2006, Banerjeeetal2013}. 
 
\begin{figure}[h]
\includegraphics[width=8.5cm]{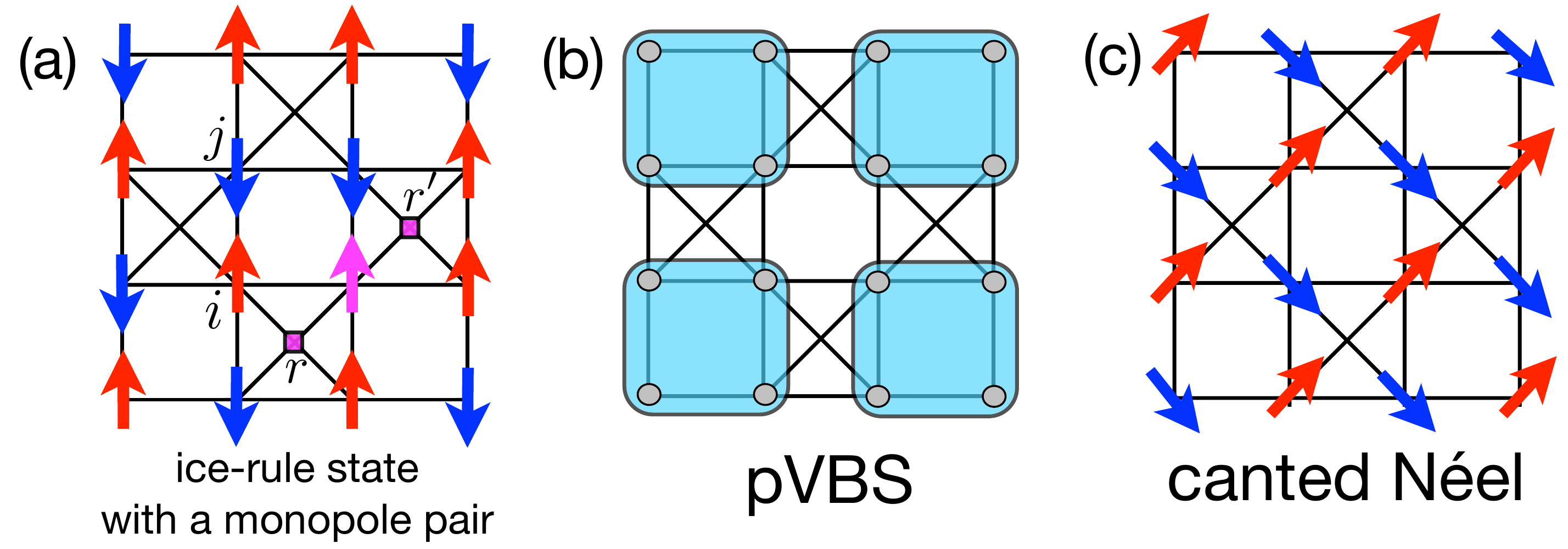} 
\caption{(a) Checkerboard lattice, showing the notation for the lattice-site indices and for the vertex indices, as well as a classical ice-rule configuration plus a monopole pair; (b-c) Sketch of the ordered ground-state phases of quantum square ice. The squares in the pVBS phase indicate resonating states of the kind $(|N\rangle + |\bar{N}\rangle)/\sqrt{2}$ (see main text for the notation).}
\label{f.sketch}
\end{figure}

  In this paper 
  we investigate the full Hamiltonian of quantum square ice realized by the transverse-field Ising model (TFIM) on a checkerboard lattice (Fig.~\ref{f.sketch}(a)). 
 Our results are based on a novel quantum Monte Carlo (QMC) scheme, which allows to efficiently update the system within the manifold of ice-rule states with diluted defects (induced by quantum fluctuations). 
  The application of a weak transverse field is confirmed to lead to a pVBS ground state (Fig.~\ref{f.sketch}(b)) via an order-by-disorder mechanism, but a stronger field drives the system through a quantum phase transition towards a (canted) N\'eel ground state (Fig.~\ref{f.sketch}(c)). Such a transition between order-by-disorder phases is found to be related to perturbation terms of 8th order in the field, going well beyond the simple fcQED description, while surprinsingly reproducing the main ingredients of abstract quantum link models recently investigated \cite{Shannonetal2004, Banerjeeetal2013}. While the N\'eel phase is seen to persist up to a sizable temperature ($\sim 1/10$ of the spin-spin coupling $J$), the pVBS melts at an exceedingly low temperature -- well below the energy scale set by the transverse field. The melting of the pVBS leads therefore to a \emph{thermal} Coulomb phase with unbound spinon excitations, whose deconfined nature is exposed by treating quantum spin ice within a gauge mean-field theory \cite{SavaryB2012}, recently introduced to describe the U(1) spin liquid of 3$d$ quantum spin ice.  We discuss the potential realization of quantum square ice in the context of atomic physics and solid-state simulators.
  
  \emph{Model.} The Hamiltonian of the TFIM on the checkerboard lattice reads
  \begin{equation}
  {\cal H} = J\sum_{\boxtimes} (\sigma^z_{\boxtimes})^2 - \Gamma \sum_i \sigma_i^x
  \label{e.Ham}
  \end{equation}
 where the first sum runs over the crossed plaquettes (vertices) of the checkerboard lattice (see Fig.~\ref{f.sketch}(a)), and $\sigma^z_{\boxtimes} = \sum_{i\in \boxtimes} \sigma_i^z$. $\sigma_i^{x(z)}$ are Pauli matrices. A Trotter-Suzuki decomposition \cite{Suzuki1993} with $M$ Trotter steps at an inverse temperature $\beta$  maps the quantum partition function of the system onto the partition function of stacked spin-ice planes with reduced couplings $J/M$, and interacting via ferromagnetic couplings of strength $J_{\tau} = -\log[\tanh(\beta \Gamma/M)]/(2\beta)$.  This mapping has the advantage that the efficient loop algorithm for spin ice \cite{NewmanB1998} can be generalized to the quantum context, where it takes the form of a \emph{membrane} algorithm: a loop of spin flips (or an open string in the presence of monopole excitations) is first created at a given imaginary time, and then propagated along the imaginary-time direction as in a 1$d$ Wolff algorithm \cite{suppmat}.  The resulting dynamics allows to explore efficiently the delicate coexistence between kinematic constraints and quantum fluctuations; the introduction of the membrane move turns out to be crucial for the correct equilibration of the system, similarly to what observed for the loop move in the classical case.  Our quantum Monte Carlo simulations are performed on $L\times L$ lattices with sizes ranging up to $L=32$. 
 
\begin{figure}[h]
\includegraphics[width=8cm]{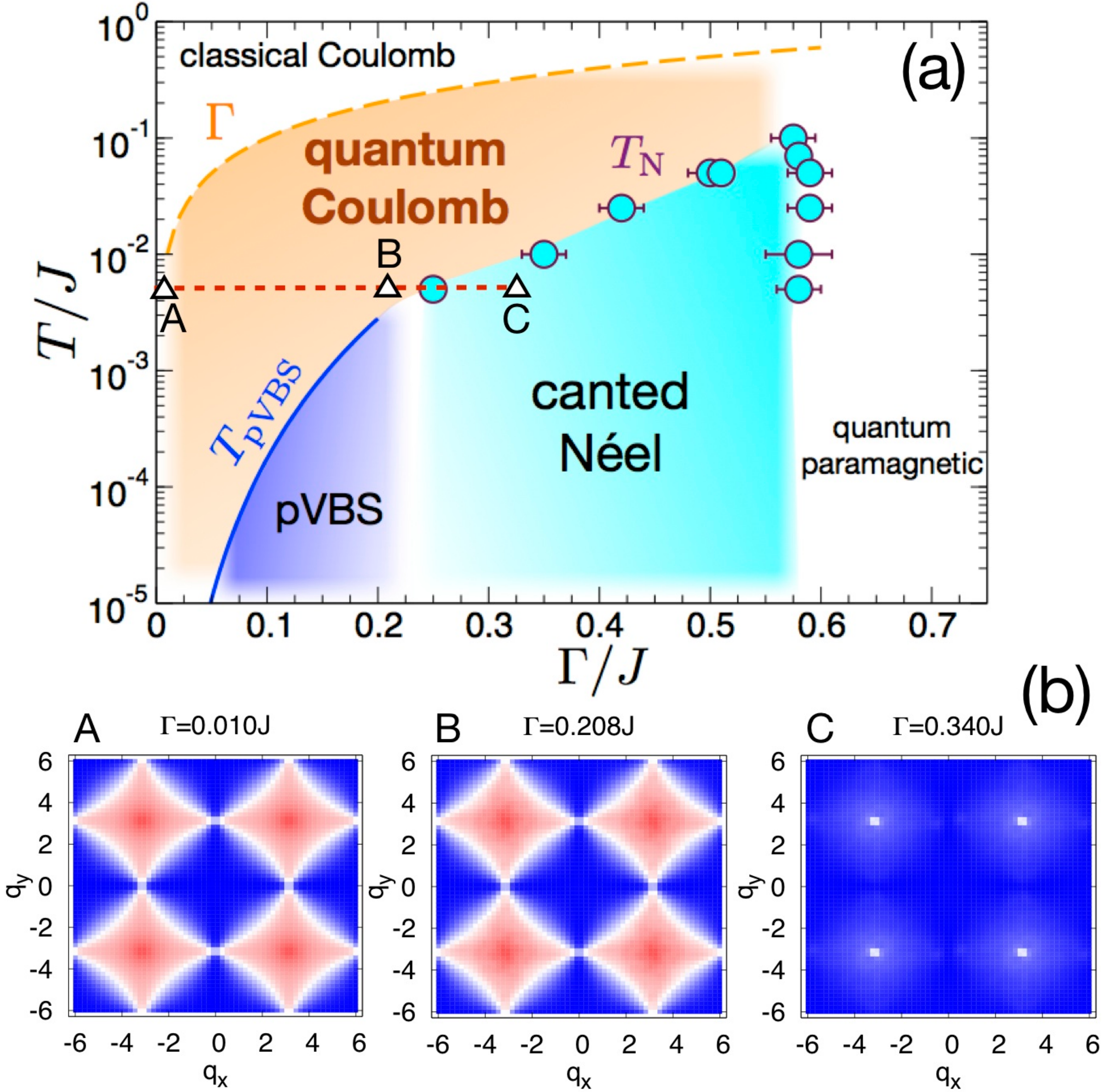} 
\caption{(a) Phase diagram of quantum square ice; boundaries of the pVBS phase ($T_{\rm pVBS}$) and of the canted N\'eel phase ($T_{\rm N}$) have been obtained as described in the main text; the dashed line marks a crossover from coherent to diffusive spinon/monopole dynamics at the energy scale set by the transverse field $\Gamma$. (b) Static structure factor for a system with $L=24$, corresponding to the $(\Gamma, T)$ parameters as indicated in panel (a).}
\label{f.PhD}
\end{figure}

 \emph{Phase diagram.} Fig.~\ref{f.PhD} shows the phase diagram of the system in the field-temperature plane. Notice the logarithmic temperature scale, emphasizing that salient features occur at very low temperatures. Upon increasing the field, the system's ground state is driven from a Coulomb phase for $\Gamma = 0$ to a pVBS phase, for $\Gamma/J \lesssim 0.25$; to a canted N\'eel phase for $0.25 \lesssim \Gamma/J \lesssim 0.55$; and finally to a quantum paramagnetic phase for $\Gamma/J \gtrsim 0.55$. The pVBS phase and N\'eel phase melt at a finite critical temperature, which has been determined as described below. 
\begin{figure}[h]
\includegraphics[width=7cm]{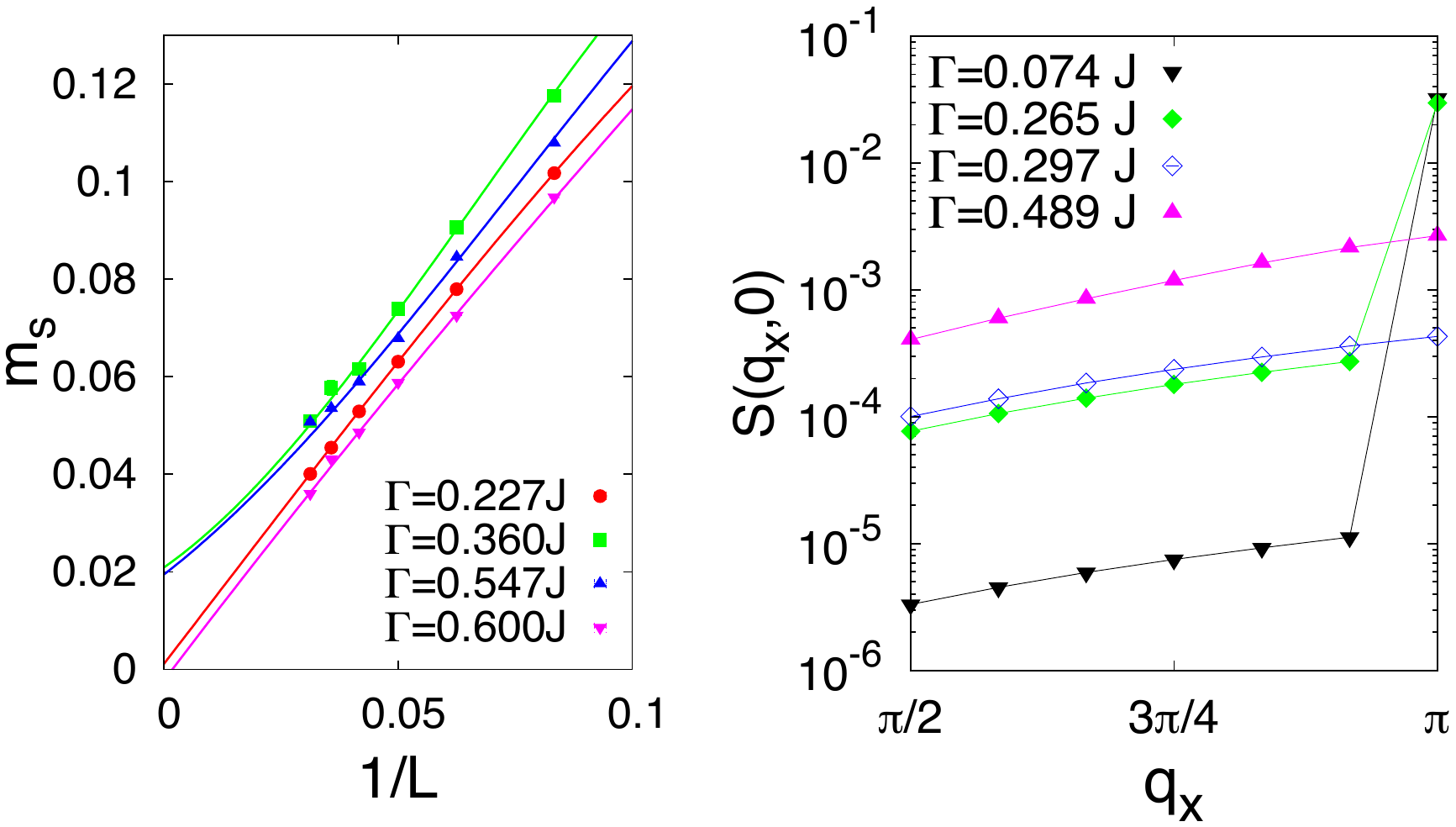}
\caption{(a) Scaling of the N\'eel order parameter at $T/J = 10^{-2}$; solid lines are fits to cubic polynomials; (b) Scans in the static structure factor at $T/J = 5\times 10^{-3}$ and $L=24$, showing the evolution of the pinch-point width.} 
\label{f.Neel}
\end{figure}

 \emph{VBS transition and fcQED.} The appearance of a pVBS phase has been proven numerically \cite{Shannonetal2004, SyljuasenC2006} for the effective Hamiltonian obtained via degenerate perturbation theory from Eq.~\eqref{e.Ham} at the lowest (4th) order in the field
 \begin{equation}
 {\cal H}^{(4)}_{\rm eff} = - K_4 \sum_{\square} F_{\square} + {\rm const.}
 \label{e.H4eff}
 \end{equation}
 corresponding to fcQED for a $S=1/2$ discrete gauge field. 
 Here the sum runs over the (uncrossed) plaquettes, and $F_{\square} = \sigma_1^{+} \sigma_2^{-} \sigma_3^{+} \sigma_4^{-} + {\rm h.c.}$ is the plaquette flip operator (the indices run counterclockwise around the plaquette). The coupling constant has value $K_4 = 20 \frac{\Gamma^4}{(2J)^3}$, where
 the factor of $20$ accounts for all the possible sequences of elementary spin flips leading to a plaquette flip, and creating either one or two monopole pairs as virtual intermediate excitations (see \cite{suppmat}). 
 The characteristic ordered structure of the pVBS state corresponds to the appearance of a staggered pattern of local resonances between a plaquette N\'eel state $|N\rangle = |\uparrow_1 \downarrow_2 \uparrow_3 \downarrow_4\rangle$  and its spin-flipped partner $|\bar{N}\rangle$ (see sketch in Fig.~\ref{f.sketch}). Such an ordered structure can be captured by the \emph{flippability}, namely the average value of the projector onto flippable (=N\'eel) plaquette states 
$ f_{\square} = \langle | N \rangle \langle N | + | \bar N \rangle \langle \bar N |\rangle = \langle F_{\square}^2\rangle $. 
Detecting directly the onset of pVBS order for the TFIM Hamiltonian of Eq.~\eqref{e.Ham} turns out to be prohibitive from the numerical point of view, given that the onset of pVBS order occurs at a temperature $T/J \sim (\Gamma/J)^4$ lying several orders of magnitude below the energy scale of the spin-spin coupling. We rather focus on the effective Hamiltonian Eq.~\eqref{e.H4eff}, and calculate its thermal phase transition to pVBS order via path-integral Monte Carlo (PIMC) \cite{suppmat} -- for such a system, the membrane algorithm is the only possible update compatible with the kinematic constraints. 
Using the crossing of the Binder cumulants for the flippability \cite{suppmat} we determine the critical temperature for the melting of the pVBS state as $T_{\rm pVBS}/J= 1.75(5) (\Gamma/J)^4 $. This estimate allows us to draw the curve shown in Fig.~\ref{f.PhD}.

\begin{figure}[h]
 \includegraphics[width=6cm]{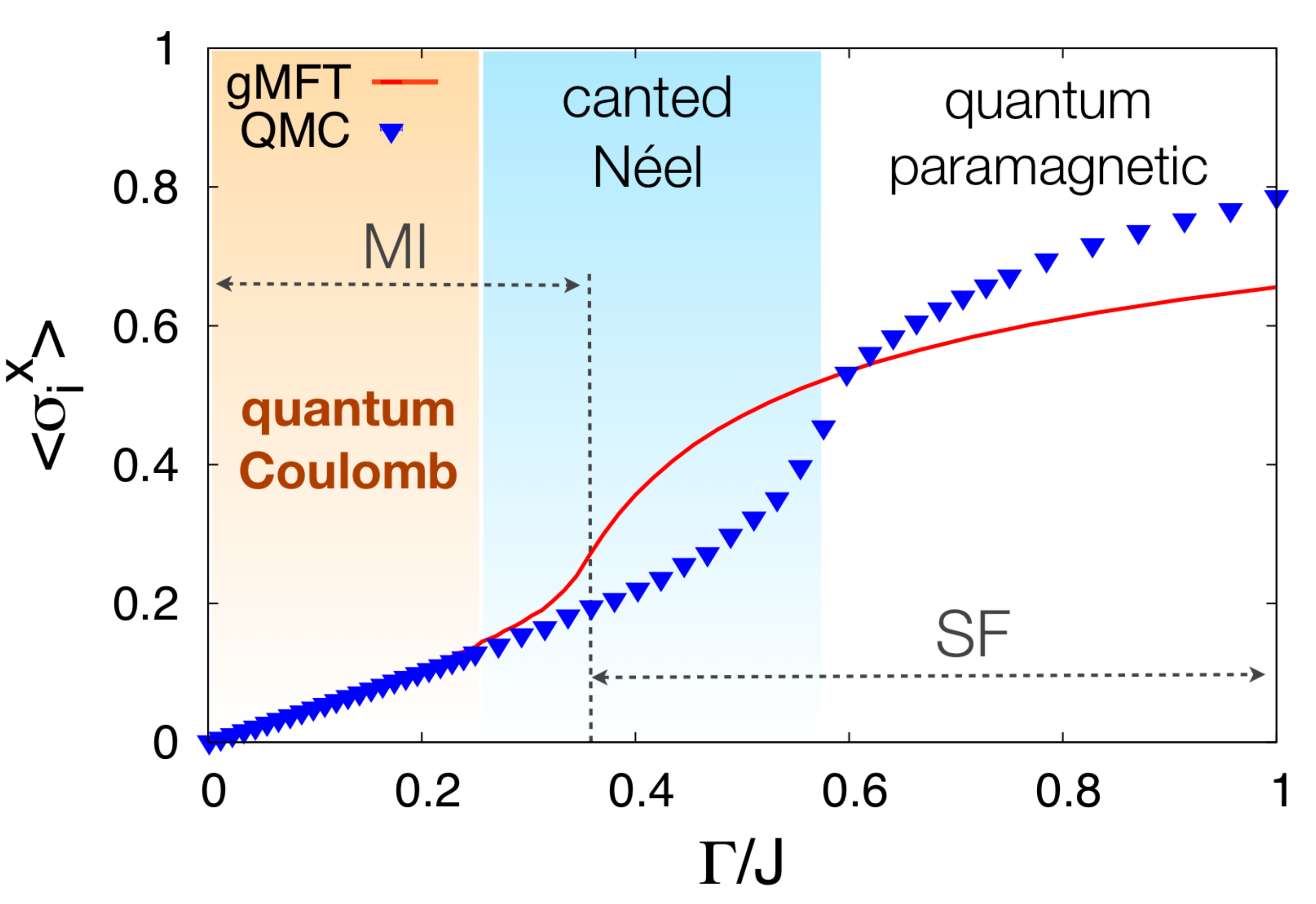}
\caption{Transverse magnetization of quantum square ice for $T/J=5\times 10^{-3}$ and $L=16$, compared with the gauge mean-field theory (gMFT) prediction. The vertical dashed line marks the transition from Mott insulator (MI) to superfluid (SF) in the corresponding quantum rotor model (see text).} 
\label{f.Sx}
\end{figure}

 \emph{N\'eel transition.} In the case of the N\'eel phase, we have considered systematic finite-size extrapolations of the staggered magnetization, estimated as $m_s = (1/L^2) \langle \left | \sum_i(-1)^i  \sigma_i^z \right| \rangle$, where $L$ is the linear size of the system. Fig.~\ref{f.Neel} shows polynomial fits to the finite-size dependence of the magnetization, exhibiting a very small ($\sim 10^{-2}$) albeit finite staggered moment in the thermodynamic limit. The upper critical field estimated via the vanishing of the order parameter is found to be consistent with the position of an inflection point in the transverse magnetization (see Fig.~\ref{f.Sx}). 

 \emph{N\'eel phase from the effective Hamiltonian.} The appearance of the N\'eel phase is a highly non-trivial order-by-disorder phenomenon, as it is associated with diagonal order induced by a purely non-diagonal operator (the transverse field term) - and, paradoxically, it appears only if the transverse field is sufficiently strong, while at weak fields the order is rather off-diagonal. One might suspect that such a phase is already present in the classical ($S\to \infty$) limit of the TFIM due to an order-by-disorder mechanism induced by thermal fluctuations; we have checked explicitly this aspect (see \cite{suppmat}) and we do not find any form of magnetic order in the classical, continuous spin version of Eq.~\eqref{e.Ham} at small but finite temperature. Moreover the N\'eel phase is not stabilized by harmonic quantum fluctuations, as verified explicitly within spin-wave theory in Ref.~\onlinecite{Henryetal2012}.    
 
 The understanding of this phase can only be gained when going beyond the lowest-order perturbative Hamiltonian of Eq.~\eqref{e.Ham}, and considering further perturbation terms. One can do so systematically following \emph{e.g.} Ref.~\cite{DPT} - see \cite{suppmat} for an extensive discussion. In general the effective Hamiltonian within degenerate perturbation theory represents a most general U(1) gauge theory of the quantum-link model type in the \emph{pure gauge} sector, namely in the absence of matter (which for quantum spin ice is represented by monopoles). The gauge symmetry of the effective Hamiltonian is not shared by the original TFIM Hamiltonian, and in particular the ground state of the TFIM Hamiltonian does contain a finite concentration of monopoles (resulting in a finite transverse magnetization -- see below for further discussion). This implies that the ground state of the TFIM will not have the same topological properties as the effective Hamiltonian (the transverse field mixes several topological sectors, while the effective Hamiltonian does not); but one expects the same symmetry breaking phenomena to be exhibited by the ground states of both Hamiltonians. To gain a quantitative understanding of the N\'eel phase, it turns out to be necessary to push the perturbative expansion up to \emph{8th} order in the magnetic field; to this order the effective Hamiltonian - obtained by considering exclusively virtual processes involving the creation/annihilation of a single monopole pair - reads:
 \begin{eqnarray}
 {\cal H}^{(8)}_{\rm eff} & =& -K_4 \sum_{\square} F_{\square} - K_6 \sum_{l\in {\cal L}_6} F_{6l} \nonumber \\
 &-&  K_8 \sum_{l\in {\cal L}_8} F_{8l} - K'_8 \sum_{\square} F_{\square}^2 + {\rm const.} 
 \label{e.H8eff}
 \end{eqnarray}
Here $F_{nl} = \sigma_1^{+} \sigma_2^{-}... \sigma_{n-1}^{+} \sigma_n^{-} + {\rm h.c.}$ is the operator flipping the spins (in alternate fashion) on a loop $l$, belonging to the family ${\cal L}_n$ of loops of length $n$. The coefficients $K_{n} = a_n \Gamma^{n}/(2J)^{n-1}$ are given explicitly in \cite{suppmat}. The last term is a purely \emph{diagonal} term, which amounts to counting the number of flippable plaquettes, and  therefore its energy is minimized by the N\'eel state, being the maximally flippable state \cite{MoessnerS2001}. Hence we can expect that the pVBS-N\'eel transition is fundamentally driven by the competition between the 4th order term and the diagonal 8th order term. Indeed an Hamiltonian comprising exclusively those two terms has been studied in Refs.~\cite{Shannonetal2004, Banerjeeetal2013} using exact diagonalization, and a transition from N\'eel to pVBS is predicted to occur for a critical ratio $\alpha = K'_8/K_4 = \alpha_c \approx 0.37$. The ratio between these two coefficients can be controlled in the TFIM via the transverse field, $\alpha = (a_8'/a_4) (\Gamma/2J)^4$. The field corresponding to $\alpha_c$ is $\Gamma/J \approx 0.64$, a value which lies reasonably close to the field range in which the N\'eel order is seen to appear in Fig.~\ref{f.PhD} (also considering that we have arbitrarily discarded from this analysis all the other terms of Eq.~\eqref{e.H8eff} beside the first and last one, as well the processes involving more than a single monopole pair). 

 \emph{Quantum Coulomb phase.} Finally, we focus on the thermally disordered phase in quantum square ice. As already mentioned in the introduction, pinch points with \emph{zero} width in the static structure factor $S({\bm q}) = (1/N) \sum_{ij} e^{i{\bm q}\cdot({\bm r}_i - {\bm r}_j)} \langle \sigma_i^z \sigma_j^z \rangle$ are a consequence of algebraic spin-spin correlations of the classical Coulomb phase of 2$d$ spin ice \cite{Youngbloodetal1980}, which are in turn a characteristic feature of the spatial correlations of the divergenceless magnetization field of square ice \cite{suppmat}. Fig.~\ref{f.PhD}(b,A-B) and Fig.~\ref{f.Neel} shows that, for weak field and low temperatures (namely for the paramagnetic phase lying immediately above the pVBS phase), pinch-point features survive in the structure factor, despite the fact that the transverse field induces a finite concentration of monopoles in the system. This suggests that the spin-spin correlation length, even if finite due to the finite temperature and the finite concentration of monopoles, remains extremely large. Indeed monopole pairs induced by the transverse field are strongly off-resonant (as $\Gamma \ll 2J$), and hence they form bound states; as a consequence they screen each other, only moderately affecting the spin-spin correlations. This interpretation is strongly corroborated when considering that the typical distance between monopole defects induced by the field can be estimated as the typical distance between two spins flipped by the transverse field, namely $l_{\Gamma} = (\langle \sigma^{x} \rangle)^{-1/2}$ - giving $l_{\Gamma} \sim 2$ when $\langle \sigma^{x} \rangle=0.2$; on the other hand we observe in Fig.~\ref{f.Neel} that the width of the pinch points is resolution-limited for system sizes up to $L=24$, and fields up to the N\'eel transition, beyond which the pinch point broadens abruptly.  Hence up to the N\'eel transition, and even for sizable system sizes, the main features of the structure factor are hardly distinguishable from those of the classical Coulomb phase at $T=0$, $\Gamma=0$ (see Fig.~\ref{f.PhD}(b,A-B)).

  The low-$T$ disordered phase for $\Gamma \lesssim 0.2 J$ preserves therefore some fundamental features of the classical Coulomb phase (at least over a finite but extremely large range), but with a fundamental difference: if enough energy is transferred to the system as to resonantly excite a monopole pair, its subsequent dynamics is not diffusive (as in the classical limit), but rather \emph{coherent}, as monopoles hop through quantum spin flips at a rate $\Gamma$ (and the temperature is $T\ll \Gamma$). One could reasonably suspect that monopole excitations are fully deconfined above the pVBS state, given that their confinement energy is of the order of the pVBS gap $\Delta_{\rm VBS} \sim \Gamma^4/(2J)^3$; hence in the temperature range $T_{\rm VBS} \leq T \lesssim \Gamma$ the elementary excitations of the system are expected to be thermally deconfined \emph{spinons} (or coherent monopoles), with a finite, albeit exceedingly large correlation length thanks to the screening of the bound monopole pairs nucleated by the transverse field. We call this regime a thermally induced \emph{quantum Coulomb phase}, whose short-range properties are identical to those of a U(1) spin liquid phase (the latter being realized strictly speaking only in 3$d$ at $T=0$ \cite{Hermeleetal2004}). In particular we expect spin-spin correlations to decay algebraically in the quantum Coulomb phase up to a length $\sim \min(l_{c}, l_{\rm th})$, where $l_c \sim \Delta^{-1}_{\rm VBS}$ is the confinement length in the pVBS phase, and $l_{\rm th} \sim \exp(2J/T)$ is the 
 the average distance between \emph{thermally} excited spinon pairs; it is easy to verify that both lengths can be extremely large in the phase in question (in particular $l_{\rm th}$ is astronomically large in the low-temperature range of the quantum Coulomb phase).      

 \emph{Gauge mean-field theory.} The picture of a quantum Coulomb phase is further corroborated by a theoretical treatment of quantum square ice based on the recently introduced gauge mean-field theory (gMFT) \cite{SavaryB2012}. The latter approach formally splits the $S=1/2$ spin degrees of freedom into a ``matter" part - the spinon field, represented by a bosonic field of integer modulus $\Phi_{\rm r} = e^{i\phi_{r}}$ -  living on the centers ${r}$ of the vertices, and a gauge part  - the $S=1/2$ spin gauge field $s^{\alpha}_{rr'}$, with $\alpha = x,y,z$ -  living on the sites of the lattice which are in between two vertices $r$, $r'$ (see Fig.~\ref{f.sketch}(a)). A mean-field decoupling of the gauge field with respect to the spinon field leads to the following Hamiltonian, 
 ${\cal H} \approx {\cal H}_{\Phi} +  {\cal H}_{s} + {\rm const.}$,  \cite{suppmat} with
 \begin{eqnarray}
 {\cal H}_{\Phi} &=& - 2\Gamma  \sum_{\langle { rr'} \rangle} \langle s_{ rr'}^{x} \rangle \cos(\phi_{ r} - \phi_{ r'}) + 4J \sum_r Q_r^2  \label{e.qrotor}\\
 {\cal H}_{s} &=& -2 \Gamma \sum_{\langle { rr'} \rangle} \langle \cos(\phi_{ r} - \phi_{ r'}) \rangle~ s_{ rr'}^x ~.
 \label{e.gauge}
 \end{eqnarray}
 Here $Q_r$ is the conjugate (charge) operator to the spinon phase, $[\phi_r, Q_r]=i$. 
In particular ${\cal H}_s$ is readily minimized by a state with $\langle s^x \rangle = 1/2$, reducing the spinon Hamiltonian ${\cal H}_{\Phi}$ to a \emph{quantum rotor Hamiltonian} on the square lattice. Within this mapping the transverse magnetization is simply related to the kinetic energy of the bosonic spinons, namely $\langle \sigma^x \rangle =  \langle  \cos(\phi_{ r} - \phi_{ r'})\rangle$. Remarkably, the quantum rotor Hamiltonian admits a numerical solution via PIMC (see \cite{suppmat} for the details), which allows us to compare quantitatively the predictions of gMFT  with the exact results coming from the PIMC simulation of quantum square ice. This comparison is made in Fig.~\ref{f.Sx}, clearly showing that $T=0$ gMFT is quantitatively accurate in the quantum Coulomb phase, while it deviates from the numerically exact results for quantum square ice precisely when the system enters the N\'eel phase. The above agreement holds despite the fact that gMFT ignores the gauge-field dynamics, and its confining effect on the matter excitations - showing that such an effect is \emph{not} at play in quantum square ice already at very low temperature ($T=5\times 10^{-3}J$ for the data in Fig.~\ref{f.Sx}). In particular, gMFT represents the low-field phase for the matter sector of quantum square ice as a bosonic \emph{Mott insulator}, with a gap corresponding to the spinon gap, and spinon pairs representing particle-hole pairs of the Mott insulator. Indeed the ground state of a Mott insulator experiences quantum nucleation of bound particle-hole pairs, corresponding to the vacuum spinon-pair fluctuations in quantum square ice (see Fig.~\ref{f.sketch}(a)).  The result of such fluctuations is a finite kinetic energy of the spinons from Eq.~\eqref{e.qrotor}, and a corresponding finite transverse magnetization. Most importantly, the elementary excitations of a bosonic Mott insulator are gapped, deconfined particle-hole pairs forming a continuum \cite{BH}. Therefore this result further corroborates the picture in which the excitation spectrum for the matter sector of the quantum Coulomb phase consists of a continuum of deconfined spinons.      

 \emph{Experimental realization.} 
  The most prominent experimental platform for the realization of quantum square ice is represented by micro-trapped ions, which naturally implement transverse-field Ising models in different planar geometries \cite{Schneideretal2012}; an alternative scheme might rely on tailored nanomagnets \cite{Khajetooriansetal2012} composed of Ising-like magnetic moments. Trapped-ion experiments typically feature dipolar interactions, while tailored nanomagnets exhibit Ruderman-Kittel-Kasuya-Yosida interactions, both possessing a long-range tail, and leading to a possible asymmetry between the nearest-neighbor and next-nearest-neighbor couplings (already considered in Ref.~\onlinecite{Henryetal2012}). The long-range interactions might destabilize the pVBS phase (while they might further stabilize the N\'eel phase), but they are expected to have a marginal impact on the quantum Coulomb phase as long as the low-$T$ symmetry breaking phase, induced by the long-range tails of the interactions, melts at a critical temperature $T_c \ll \Gamma$.  This suggests that atomic physics or solid-state quantum simulators offer promising platforms for the implementation of fundamental phenomena of lattice gauge theories (such as confinement/deconfinement transitions) - similar ideas are currently the subject of intense theoretical investigations in the context of neutral atoms \cite{Banerjeeetal2012, Zoharetal2012, Tagliacozzoetal2013}, specifically aimed at the realization of U(1) quantum link models \cite{Banerjeeetal2012, Tagliacozzoetal2013,Banerjeeetal2013}.

We acknowledge fruitful discussions with P. Holdsworth and F. B\`egue. All calculations have been performed on the computer cluster at the PSMN (ENS Lyon), whose support we gratefully acknowledge.

\newpage 

\setcounter{page}{1}
\setcounter{figure}{0}
\setcounter{equation}{0}

\section{Supplementary Material to "Order-by-disorder and quantum Coulomb phase in quantum square ice"}

\subsection{Mapping between the 16-vertex model and the Ising model on the checkerboard lattice}

\begin{figure}[h!]
 \includegraphics[width=9cm]{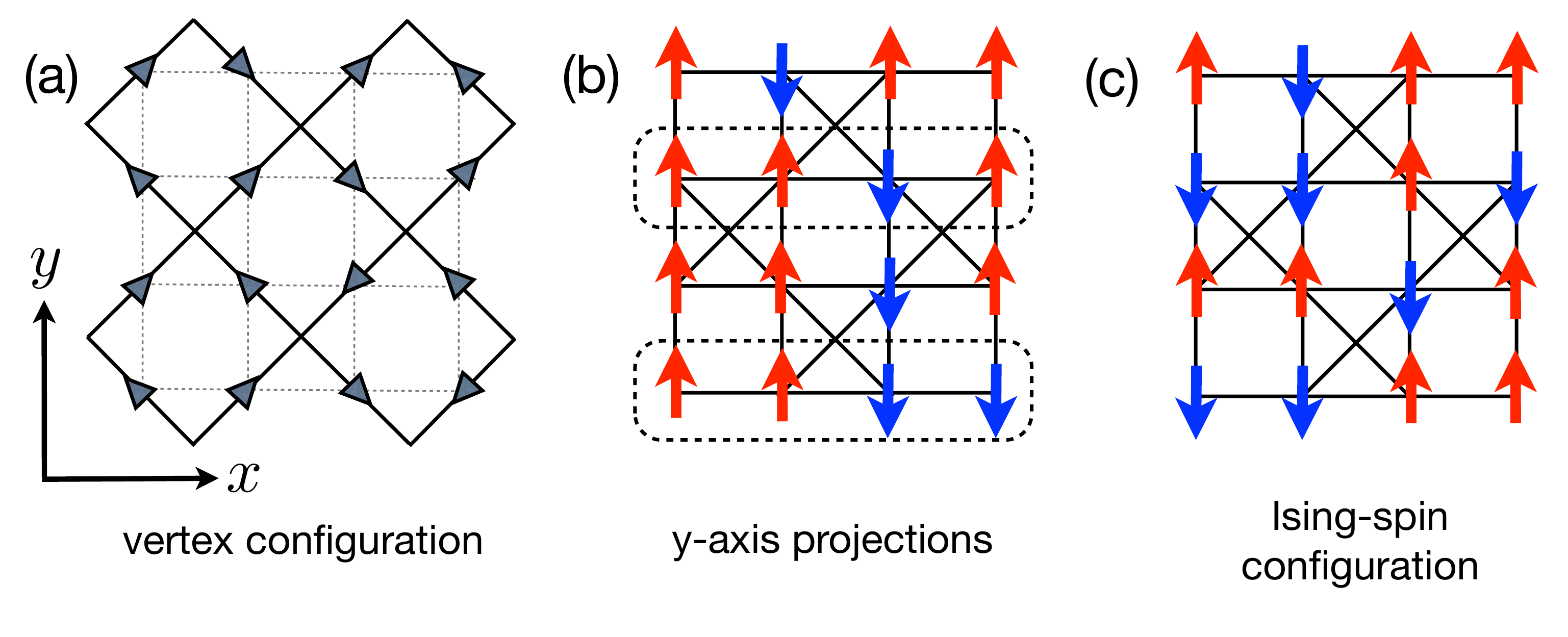}
\caption{Mapping between a vertex configuration and an Ising-spin configuration - see description in the text.} 
\label{f.map}
\end{figure}

In order to connect the observables of the antiferromagnetic Ising model on the checkerboard lattice with those of the 16-vertex model it is useful to recall the mapping which leads from the latter model to the former. Fig.~\ref{f.map} illustrates such a mapping; starting from a 6-vertex configuration (Fig.~\ref{f.map}(a)), one maps the sign of the projections of the arrows along, \emph{e.g.}, the $y$-axis onto Ising spins (pointing up for a positive projection and down otherwise - Fig.~\ref{f.map}(b)). Flipping the Ising spins of every other row (Fig.~\ref{f.map}(c)), gives zero (Ising-spin) magnetization on each vertex if the corresponding vertex configuration is a 6-vertex one obeying the 2-in/2-out ice rule (a similar mapping is obtained by flipping every other column). In particular ice-rule vertices having counterpropagating arrows on parallel bonds are mapped onto N\'eel vertices for the Ising spins, while ice-rule vertices with copropagating arrows on parallel bonds are mapped onto collinear vertices.

 The asymptotic correlation function for the $y$ spin components of the 6-vertex model has been calculated exactly in Ref.~\onlinecite{sYoungbloodetal1980}. Introducing the spin-flip of every other row, this translates into the following behavior for the Ising-spin correlation function
 \begin{equation}
 \langle \sigma_i^z \sigma_j^z \rangle \sim (-1)^{y} \frac{ x^2 - y^2}{(x^2 + y^2)^2} 
 \end{equation}
 where $x = x_i - x_j$ and $y = y_i - y_j$. The corresponding static structure factor features a pinch point around ${\bm q} = (0,\pi)$ as 
 \cite{sYoungbloodetal1980}
 \begin{equation}
 S(h,\pi+k) \sim \frac{h^2}{h^2+k^2} 
 \end{equation}
for $h, k \ll \pi$. On the other hand, the ensemble of ice-rule states is \emph{invariant} under all operations mapping ice-rule states onto ice-rule states; one such operation is the mirror reflection around the (1,1) axis, which produces a mirror pinch point around ${\bm q} = (\pi,0)$, as shown in Fig.~\ref{f.PhD} of the main text. 

 Another important exact result for the 6-vertex model due to Sutherland \cite{Sutherland1968} is that the correlation function between parallel arrows has a staggered part, decaying as $r^{-2}$, beside the non-oscillating part leading to the pinch point. When mapping to the Ising spins, this implies that the spin-spin correlation function among spins on, \emph{e.g.} the A sublattice of the square lattice (underlying the checkerboard lattice) has the simple form $\langle \sigma^z_{i \in A} \sigma^z_{j \in A} \rangle \sim r^{-2}$. This would imply a logarithmically divergent peak in the static structure factor for ${\bm q}=0$, and at the equivalent points ${\bm q} = (\pm \pi, \pm \pi)$ and 
 ${\bm q} = (\pm \pi, \mp \pi)$. In fact when considering the whole static structure factor
 \begin{equation}
 S({\bm q}) = \sum_{i,j \in A} e^{i {\bm q}\cdot({\bm r}_i - {\bm r}_j)} \left\langle \sigma^z_i \sigma^z_j \left( 1+e^{i q_x} \sigma^z_i \sigma^z_{i'} \right)
 \left( 1+e^{i q_x} \sigma^z_j \sigma^z_{j'} \right) \right\rangle
 \end{equation}
 (where $i'(j') \in B$ is the nearest neighbor to $i (j)$ in the same unit cell), one observes that the unit-cell form factors $1+e^{i q_x} \sigma^z_{i(j)} \sigma^z_{i'(j')}$ suppress the peak at ${\bm q} = 0$, given that ice-rule states typically display an antiferromagnetic configuration ($\sigma_i^z \sigma_{i'}^z = -1$) on the unit cell -- 4 out of 6 ice-rule states verify this property. Hence the static structure factor displays a logarithmically divergent peak only for ${\bm q} = (\pi,\pi)$ and equivalent points.        


\subsection{The membrane algorithm for quantum spin ice}

Here we describe the extension of the loop algorithm, of crucial importance for the simulation of classical spin ice \cite{sNewmanB1998, MelkoG2004}, to the case of quantum spin ice. The Trotter-Suzuki (TS) mapping \cite{sSuzuki1993} of the quantum partition function of a transverse-field Ising model (TFIM) allows to map the model in question onto a $(d+1)$-dimensional classical Ising model. If $M$ Trotter steps are used in the TS decomposition, the partition function takes the form 
${\cal Z} \approx \int {\cal D}(\{\sigma_{i,k}\}) \exp[-\beta S_{\rm eff}]$, involving the effective action 
\begin{equation}
S_{\rm eff}(\{\sigma_{i,k}\}) = \frac{J}{M} \sum_{k=1}^M \sum_{\boxtimes} \left(\sigma_{\boxtimes,k}\right)^2 - J_{\tau} \sum_{i,k} \sigma_{i,k} \sigma_{i,k+1}  
\label{e.Seff}
\end{equation}
where $\sigma_{i,k}$ is the Ising variable at lattice site $i$ and Trotter (imaginary-time) step $k$, $\sigma_{\boxtimes,k} = \sum_{i\in\boxtimes}  \sigma_{i,k}$, and 
$J_{\tau} = |\log(\tanh{\epsilon})|/2\beta$ with $\epsilon = \beta \Gamma / M < 1$ by construction.  
Hence quantum square ice is TS-mapped onto stacked, classical spin-ice layers interacting ferromagnetically. 

The membrane algorithm consists then in building a loop (as in the loop algorithm) in a spin-ice layer at imaginary time step $k$, or an open string in the presence of defect vertices - the latter being induced either by quantum or by thermal fluctuations. An open string is built so as to touch at most one defect vertex containing a single monopole, and if so, the defect vertex lies necessarily on one of the string end points - hence a string which does not touch any defect vertex closes on itself forming a loop. In the absence of the $J_{\tau}$ couplings, the loop (or open string) can be flipped at zero energy cost - in particular, a flipped open string has the effect of ``teleporting" the defect vertex from one of its ends to the opposite one. Yet, in the presence of the $J_{\tau}$ couplings, the loop/string flip will cause an energy variation; for $\epsilon \ll 1$ (which is the fundamental requirement for the TS approximation to be accurate), the ferromagnetic couplings are extremely strong (diverging like $|\log(\epsilon)|$) and hence one can reasonably expect that the energy variation induced by the loop/string flip is best cured by proposing an identical flip on the two neighboring layers at imaginary time steps $k-1$ and $k+1$. This amounts then to grow the loop/string into the imaginary time dimension, namely into a membrane. The membrane grows \emph{e.g.} to the $(k+1)$-th layer with a probability 
\begin{equation}
P(k \to k+1) = 1-\exp\left[\min\left(0, -2\beta J_{\tau} \sum_{i\in {\cal L}} \sigma_{i,k} \sigma_{i,k+1}\right)\right]  
\end{equation}   
and ${\cal L}$ is the loop/string. The above probability $P$ corresponds to the cluster growth probability for the Wolff algorithm \cite{sWolff1989}, performed along the imaginary-time dimension. 

Once the membrane has been grown, the flip of its spins is not automatic, because one still has to consider the energy change on the bonds connecting the membrane spins and those on its contour in real space. Hence the membrane is flipped with probability
\begin{equation}
P_{\rm flip} = \min\left[ 1, \exp\left(- \frac{2\beta J}{M} \sum_{(i,k)\in {\cal M}} ~{\sum_{j \in {\cal N}_{i}}}'  \sigma_{i,k} \sigma_{j,k} \right) \right]
\label{e.flip}
 \end{equation}  
where ${\cal M}$ is the ensemble of membrane spins, ${\cal N}_i$ represents the set of lattice sites neighboring the site $i$, and the primed sum indicates that one has to exclude the sites belonging to the membrane.  
The probability $P_{\rm flip}$ has value 1 in the classical limit $\epsilon \to 0$, $J_{\tau} \to \infty$, in which all the layers display the same configuration, and hence a microcanonical loop/string on a layer is equally microcanonical on every other layer  - obviously the membrane length in the imaginary-time dimension is $M$.  
For a finite transverse field, on the other hand, the flip probability will be typically reduced due to the presence of discontinuities in the imaginary-time propagation - associated with defect vertices (namely monopoles) appearing in isolated layers. A na\"ive estimate of the scaling of the membrane flip probability gives $P_{\rm flip} \sim \exp[-\beta (J/M) N_{\cal M} n_m]$, where $n_m$ is the density of (free) monopoles in the system, and $N_{\cal M}$ is the number of spins belonging to the membrane. Such a scaling would imply that the probability is inevitably suppressed exponentially as the temperature is decreased. Yet we will argue in the following that this is not the case. 

We observe that, if membranes are built from long (namely self-intersecting) loops, then  $N_{\cal M}/M = l_{\cal L} \sim L^{5/3}$, associated with the known scaling of the long-loop length $l_{\cal L}$ with system size $L$ \cite{NewmanB1998}. On the other hand, the length of short loops does not scale with system size, so that ${\cal M}/M \sim O(1)$ \cite{shortloops}. Hence the choice of short loops as pedestals of the membranes boosts the acceptance rate. Moreover, at very low temperatures, $\beta J \gg 1$, the thermal monopole density $n_m$ is exponentially suppressed, while the monopoles induced by quantum fluctuations are bound, as discussed in the main text. Hence their effect on the suppression of the flip probability is not as simple as their density $n_m$ appearing in the previous scaling formula. 

In particular a simple estimate (coming from perturbation theory) of the typical size of a bound monopole pair gives $l_{\rm pair} \sim |\log(\Gamma/(2J))|^{-1}$. We can therefore imagine that the flip probability of a membrane ${\cal M}$ built upon a loop/string ${\cal L}$ will be affected by bound monopole pairs only if such monopole pairs cross the loop/string, hence if they fall within a region of size $l_{\cal L} \times l_{\rm pair}$. The density of monopole pairs in the $(d+1)$ dimensional sample can be estimated as $n_{\rm pairs} \sim \langle \sigma^x \rangle/M$ (as each spin flip contributing to the transverse magnetization corresponds to a monopole pair). 
This means that the exponential suppression of $P_{\rm flip}$ due to bound monopole pairs can be estimated as $P_{\rm flip} \sim \exp(-\beta J l_{\cal L}  l_{\rm pair} \langle \sigma^x \rangle/M)$. Working at a fixed length of the Trotter step $\delta \tau = \beta/M$, and if $l_{\cal L} \sim O(1)$ (using short loops), we find that the membrane flip probability is \emph{not} reduced when lowering the temperature, and that the exponent is of $O(1)$, implying a sizable acceptance rate (in fact quite large if $\delta\tau J \ll 1$). This conclusion is corroborated by the numerically observed temperature scaling of the acceptance rate for the membrane flip \cite{LPprep}. 

Given the very strong correlations between neighboring layers, we observe that the membrane typically extends over a significant fraction of the imaginary-time dimension. As the linear size of Wolff clusters is related to the correlation length of the system \cite{sWolff1989}, we deduce that the imaginary-time correlation length is very large, as the system has a very small spectral gap, associated with the quantum lifting of the degeneracy between the ice-rule states. Hence the membrane moves has the important virtue of producing very low-energy moves which allow to explore efficiently the very dense energy spectrum at low energy -- similarly to the loop algorithm for classical spin ice, which allows to explore microcanonically the whole ice-rule manifold. 
    
Our PIMC simulations of quantum square ice are typically performed with a Trotter parameter $\epsilon = 10^{-2}$, guaranteeing a very small Trotter error on the observables of interest (transverse magnetization, static structure factor). To ensure ergodicity, we supplement the membrane algorithm with Metropolis single-spin flips, as well as with traditional Wolff clusters on the effective $(d+1)$-dimensional Ising model of Eq.~\ref{e.Seff}. 
A Monte Carlo step is composed of $L^2/4$ short loop membrane moves and $\sqrt{L}$ long loop membrane moves, as well as of $\sqrt{M}$ Wolff cluster moves and $L^2M$ single Metropolis spin flips. Our simulation typically contains $4\times 10^4$ thermalization steps and $10^4-10^6$ measurement steps.

\subsection{Classical limit of quantum square ice}

\begin{figure}[h]
		\includegraphics[width=9cm]{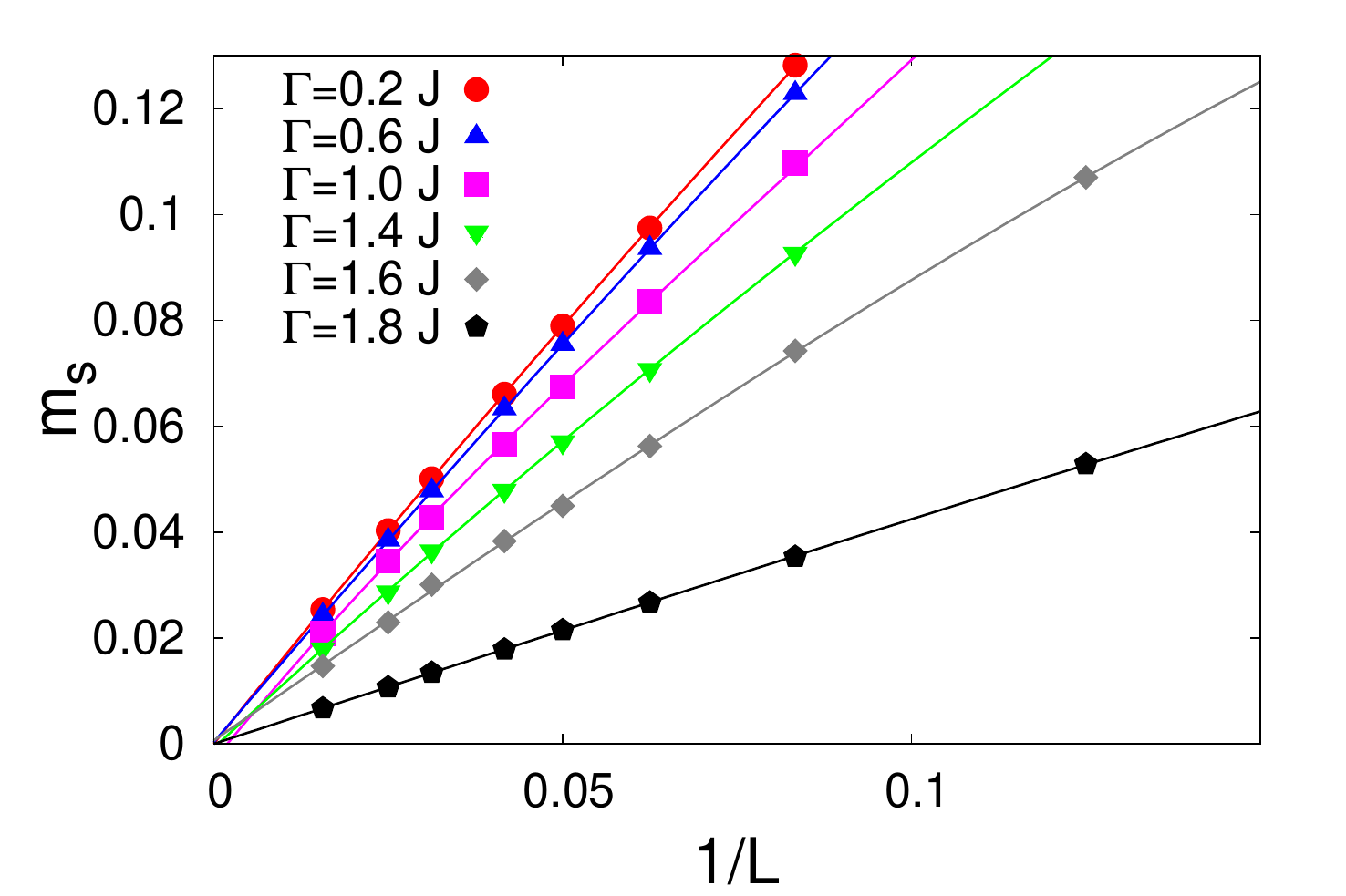} 
	\caption{Finite-size scaling of the N\'eel order parameter in the continuous spin ($S\to\infty$) limit for different values of $\Gamma$. 
		The order parameter extrapolates to 0 for the entire range of transverse field values - solid lines are fits to cubic polynomials.}
	\label{f.MNc}
	\end{figure}

Order-by-disorder phenomena in frustrated magnets can be driven either by quantum fluctuations or by thermal fluctuations - noticeable examples of the second case are \emph{e.g.} the $J_1-J_2$ antiferromagnet \cite{Henley1989} and the Kagom\'e antiferromagnet \cite{ChernM2013}. One might therefore suspect that the classically ordered phase found in quantum square ice, namely the canted N\'eel phase, is actually stabilized by thermal and not by quantum fluctuations. 
In order to check that the N\'eel ordering of quantum square ice is a purely quantum effect, we performed a classical MC simulation
 of continuous spin ($S\to\infty$) square ice in a transverse field \cite{sHenryetal2012}. The Hamiltonian reads
   \begin{equation}
  {\cal H} = J\sum_{\boxtimes} (\sum_{i\in \boxtimes} S_i^z)^2 - \Gamma \sum_i S_i^x
  \label{e.HamCl}
  \end{equation}
 Here ${\bm S}_i$ is a classical 3-dimensional vector of unit norm. We used Metropolis updates completed with generalized short- and long-loop moves.
 The loop algorithm for Ising spin ice \cite{sNewmanB1998} is generalized to the case of continuous spins in the following manner: a loop is built as for Ising spins, using the sign of the $z$ component as effective Ising spin variable; the loop flip is not microcanonical for continuous spins, and it is then accepted/rejected with Metropolis probability $P = \min[1,\exp(-\beta \Delta E)]$ where $\Delta E$ is the energy variation.   
We find no N\'eel ordering throughout the range of transverse field magnitude for which the $z$ component retains a finite value ($\Gamma \in \left[0,2J\right]$),
as it can be inferred from the finite-size scaling of the order parameter shown in Fig.~\ref{f.MNc}. Here the order parameter is estimated as $m_s^2 = (1/L^4) \sum_{ij} (-1)^{i+j} \langle S_i^z S_j^z \rangle$. 

\subsection{Path-integral Monte Carlo for frustrated compact QED}

	We have applied the membrane algorithm described above to the study of the ordering transition of the fcQED of Eq.~\ref{e.H4eff} - a detailed description will be reported elsewhere \cite{LPprep}. The transition of fcQED has apparently eluded previous numerical investigations \cite{fcQED} due to the difficulty in sampling different topological sectors of ice-rule states. The membrane algorithm guarantees on the other hand an efficient  sampling of the various topological sectors for sufficiently high temperatures and moderate system sizes.
	 
\begin{figure}[h]
		\includegraphics[width=9cm]{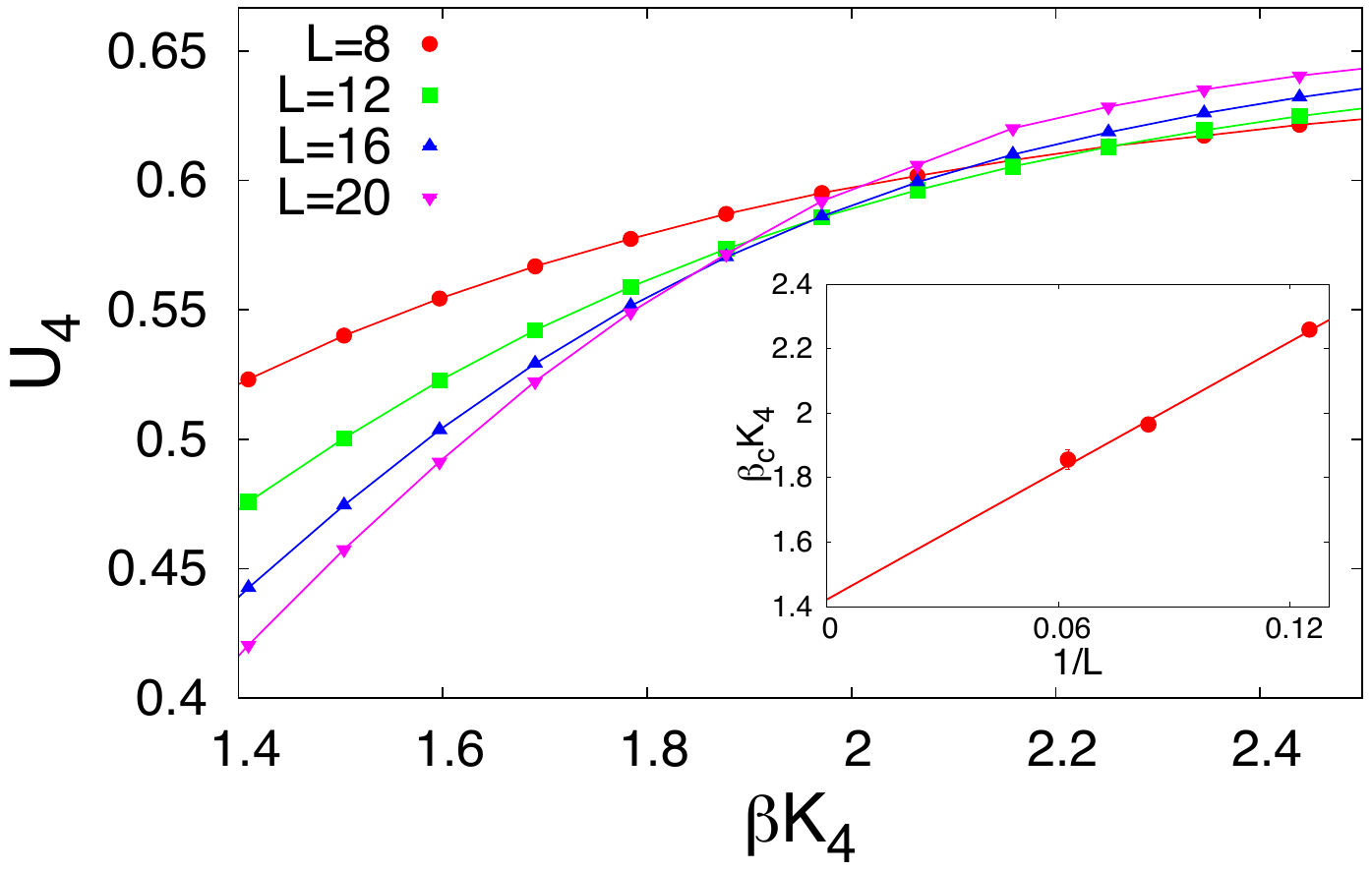} 
	\caption{Flippability Binder cumulant of fcQED for different system sizes. The curves for sizes $L$ and $L+4$ cross for $\beta= \beta_c(L)$. (\emph{inset}) Scaling of $\beta_c(L)$ with
	respect to $1/L$. }
	\label{f.Binder}
	\end{figure}

The order parameter for the pVBS phase is the staggered flippability 
	\begin{equation}
		m_{\rm pVBS}^2 = (L/2)^{-4} \sum_{\square,\square'} (-1)^{\square+\square'} \langle f_{\square}f_{\square'} \rangle
	\end{equation}
	We evaluate the critical inverse temperature $K_c$ through the calculation of the Binder cumulant
	$U_4=1-\langle m_{\rm pVBS}^4\rangle/(3\langle m_{\rm pVBS}^2\rangle^2)$ for different system sizes - shown in Fig.~\ref{f.Binder}. The crossing of the curves for
	system sizes $L$ and $L+4$ occurs at $\beta_c(L)$. We linearly extrapolate this value to $L\to\infty$ to obtain the transition
	temperature in the thermodynamic limit. The result of the extrapolation gives $\beta_c K_4=1.42(5)$, which corresponds to a transition temperature $T_c=0.70(2)K_4$.
	
\subsection{Degenerate perturbation theory for quantum square ice}

We extract the effective Hamiltonian up to 8th order in degenerate perturbation theory in the transverse field of Eq.~\eqref{e.Ham} via the resolvent method \cite{Kato,Klein}.
Using the notations ${\cal H}_0=J\sum_{\boxtimes} (\sigma^z_{\boxtimes})^2$ and $V=-\sum_i \sigma_i^x$,
the effective Hamiltonian reads
\begin{equation}
	{\cal H}_{\rm eff}=-\sum_{n=1}^{\infty}\Gamma^nP_0\left(V\dfrac{1-P_0}{{\cal H}_0}\right)^{n-1}VP_0
	\label{e.Heff}
\end{equation}
with $P_0$ the projector onto the ground-state manifold of ${\cal H}_0$ (ice-rule states). The factors $(1-P_0)/{\cal H}_0$ are sensitive to the number of virtual monopole pairs created in the intermediate configurations at the energy cost of $\Delta=2J$ per pair.
A term of order $n$ contains $n$ $\sigma_i^x$ operators, corresponding to the flip of at most $n$ spins (it can be less than $n$ because
some of spins might be flipped multiple times). 

The general form of the effective Hamiltonian in terms of projectors contains several terms which seemingly lead to super-extensive
contributions to the energy. Those terms must cancel out to recover an extensive effective Hamiltonian - we checked explicitly this aspect up to fourth order; as for higher order, we simply discard the non-extensive terms. Moreover all terms with $n$ odd necessarily vanish, as they do not conserve the vanishing magnetization of the ice-rule states. 

The off-diagonal terms in Eq.~\eqref{e.Heff} come from the flip of closed loops of spins (of even number) of alternating orientations; such flip preserves the constraint of zero magnetization on each vertex, connecting therefore different ice configurations. 
The effective Hamiltonian can then be rewritten as 
\begin{equation}
	{\cal H}_{\rm eff}=-\Delta\sum_{n=4,6,8,...}^{\infty}\left(\dfrac{\Gamma}{\Delta}\right)^n\sum_{l \in {\cal L}_n}a_{nl} F_{nl}~.
\end{equation}
Here the loop index $l$ is summed over all loops ${\cal L}_n$ of length $n$. The factors $a_{nl}$ take into account two aspects: 1) the number of sequences of elementary spin flips leading to the flip of the loop $l$ of length $n$; 2) the number of intermediate monopole pairs created in the process. In particular the $a_{nl}$ coefficients admit the following decomposition: 
\begin{equation}
a_{nl} = g^{(1)}_{nl} + \sum_{q=1}^{n-2} \frac{g^{(2q)}_{nl}}{2^q} + \sum_{q,p, q+p\leq n-1} \frac{g^{(2q,3p)}_{nl}}{2^q 3^p} + ...
\end{equation} 
where $g^{(1)}_{nl}$ is the multiplicity of spin-flip sequences leading to the virtual creation of a single monopole pair; $g^{(2q)}_{nl}$ is the multiplicity of spin-flip sequences involving the creation of two monopole pairs for $q$ configurations out of the $n-1$ virtual intermediate ones;   $g^{(2q,3p)}_{nl}$ is the multiplicity of spin flip sequences involving the creation of two monopole pairs during $q$ steps and three monopole pairs during $p$ steps, etc. 
 It is apparent that the enumeration of all processes (especially those of higher order in the number of virtual monopole pairs), represents an increasingly hard problem when going up in perturbation order. For the sake of simplicity we restrict our calculations to the one-monopole-pair term $g^{(1)}_{nl}$ only. This restriction leads then to the effective Hamiltonian Eq.~\eqref{e.H8eff} of the main text, with the following coefficients
 \begin{equation}
 K_4 = 8; ~~~~ K_6 = 96; ~~~~~ K_8 = 512; ~~~~~~ K'_8 = 288;~~~...
 \end{equation}
In particular the coefficient $K'_8$ multiplies a diagonal term, coming from the forward and backward flip of the same (flippable) plaquette, and therefore simply counting the number of flippable plaquettes. 

In the case of 4-th order term it is easy to account for all processes (involving up to two monopole pairs); this gives $a_{4l}= 20$, which we use for the exact estimate of the coefficient $K_4$ entering the Hamiltonian of fcQED.


\subsection{Gauge mean-field theory for quantum square ice}

 Gauge mean-field theory (gMFT), as introduced in Ref.~\cite{sSavaryB2012}, consists generically of a mean-field decoupling between the matter field and the gauge field in a gauge theory. In the case of quantum spin ice, one can identify an emergent lattice gauge theory description of the system in which the gauge field is essentially represented by the off-diagonal Hamiltonian terms leading to quantum fluctuations between ice-rule states, while the matter field is represented by the monopole excitations associated with the diagonal part of the Hamiltonian. Formally the gauge and matter field are not distinct mathematical objects, but they are in fact associated with different components of the same lattice spin field. In order to recover a description of spin ice in terms of a standard lattice gauge theory, it is then necessary to artificially enlarge the Hilbert space of spin variables, in order to accommodate a properly defined matter field in the system. This is done by the following redefinition of the spin operators 
 \begin{equation}
 \sigma^{+}_{rr'} \to \Phi_r^{\dagger} s^+_{rr'} \Phi_{r'} ~~~~~~  \sigma^z_{rr'} \to 2 s^z_{rr'}~.
 \label{e.gMFT}
 \end{equation}  
 Here $s_{rr'}^{\alpha}$ is a spin $S=1/2$ field (acting as the \emph{gauge} field), living on the sites of the checkerboard lattice, which represent the bonds between sites $r$ and $r'$ of the vertex lattice (see Fig.~\ref{f.sketch}(a) of the main text). The \emph{matter} field $\Phi_r$ is a bosonic field, $[\Phi_r, \Phi_r'^{\dagger}] = \delta_{rr'}$ living on the vertex lattice; it is chosen to be of unit amplitude, $\Phi_r = e^{i\phi_r}$, where $\phi_r$ is a phase operator canonically conjugated to a charge operator $Q_r$, $[\phi_r, Q_r] = i$; this choice preserves the values of the matrix elements of the spin operators. Nonetheless the newly defined spin operators of Eq.~\eqref{e.gMFT} act on a larger Hilbert space, which is infinite-dimensional (as $Q_r$ takes integer values from $-\infty$ to $+\infty$). In fact the bosonic field represents the monopole/spinon field if one enforces the constraint
 \begin{equation}
 Q_r = \frac{(-1)^r}{2} \sum_{r' {\rm(n.n.)} r} \sigma^z_{rr'} 
 \end{equation}
 where the sum runs over the vertices which are nearest neighbors of the one at position $r$ (namely on the spins contained in the vertex in question). 
 In this case $Q_r = 0, \pm 1, \pm 2$. This constraint will not be explicitly implemented in the following, but it will emerge dynamically in the relevant range of validity of the theory. 
 
 For the TFIM, the Hamiltonian acting on the enlarged Hilbert space takes the simple form 
 \begin{equation}
 {\cal H} \to 4J \sum_r (Q_r)^2 - 2\Gamma \sum_{\langle rr' \rangle} \left(\Phi_r^{\dagger} s^+_{rr'} \Phi_{r'} + {\rm h.c.}\right)~.
 \end{equation}
 The gMFT approach consists then in the mean-field decoupling 
 $$\Phi_r^{\dagger} s^+_{rr'} \Phi_{r'} \to  s^+_{rr'}  \langle \Phi_r^{\dagger}  \Phi_{r'} \rangle + 
 \langle s^+_{rr'} \rangle  \Phi_r^{\dagger}  \Phi_{r'} - \langle s^+_{rr'} \rangle  \langle \Phi_r^{\dagger}  \Phi_{r'} \rangle $$
 which leads to the Hamiltonian decomposition ${\cal H} \approx {\cal H}_{\Phi} +  {\cal H}_{s} + {\rm const.}$ as in Eqs.~\eqref{e.qrotor}-\eqref{e.gauge} of the main text. 
 
 The mean-field decoupling between the gauge field and the matter field necessarily implies that the gauge theory is described in its \emph{deconfined} phase - indeed the matter field only sees a uniform, mean-field gauge field $\langle s^x \rangle$, which is not confining. Hence such a decoupling can be applied exclusively to the thermally induced quantum Coulomb phase. Moreover the mean-field decoupling provides a featureless description of the spin gauge field, and it cannot describe the nature of the excitations in the pure gauge sector of the theory (namely the photon). On the other hand the matter sector of the theory has a non-trivial description in terms of a quantum rotor model ${\cal H}_{\Phi}$. If we interpret $Q_r = n_r - \bar{n}$ as the deviation from an average, integer density ${\bar n} \gg 1$, we see that the monopole pairs represent particle-hole excitations of a Bose fluid living on the lattice of vertices. Such a fluid is in a Mott insulator phase for $\Gamma \ll 4J$ (which is the domain of applicability of gMFT to our model): in this phase particle-hole fluctuations are suppressed, so that configurations with $|Q_r| > 2$ are energetically excluded without the need to enforce explicitly the corresponding constraint.  
 
 We solve the quantum rotor model on a square lattice using path-integral Monte Carlo, as described in Ref.~\cite{sWallinetal1991}. In particular our simulation aims at the ground-state kinetic energy $\langle \cos(\phi_i - \phi_j) \rangle $ - we observe that, for a system with $L=10$, $\beta\Gamma = 10$ and $4 \beta J/M = 10^{-2}$, thermal, finite-size and Trotter-approximation effects are all essentially removed. 
 The data shown in Fig.~\ref{f.Sx} have been obtained with the latter parameters.

\end{document}